\documentclass[prd,amsfonts,showpacs,
longbibliography,nofootinbib,amsmath,amssymb]{revtex4}
\usepackage{bm,graphics,graphicx,epsfig,soul,latexsym,hyperref,mathrsfs,multirow}

\newcommand{\ui}{\mathrm{i}}
\newcommand{\ud}{\mathrm{d}}
\usepackage[english]{babel}

\usepackage[usenames]{color}

\definecolor{darkgreen}{rgb}{0,0.5,0}

\hypersetup{
    bookmarks=true,         
    unicode=false,          
    pdftoolbar=true,        
    pdfmenubar=true,        
    pdffitwindow=false,     
    pdfstartview={FitH},    
    pdftitle={My title},    
    pdfauthor={Author},     
    pdfsubject={Subject},   
    pdfcreator={Creator},   
    pdfproducer={Producer}, 
    pdfkeywords={keyword1} {key2} {key3}, 
    pdfnewwindow=true,      
    colorlinks=true,       
    linkcolor=red,          
    citecolor=cyan,        
    filecolor=magenta,      
    urlcolor=darkgreen,           
    linktocpage=true
}

\allowdisplaybreaks

\begin{document}

\title{Ready-to-use post-Newtonian gravitational waveforms for binary black holes 
with non-precessing spins: An update}

\author{Chandra Kant Mishra}\email{chandra@icts.res.in}
\affiliation{International Centre for Theoretical Sciences-Tata Institute of
Fundamental Research, Bangalore, 560089, India}

\author{Aditya Kela}\email{akela@thp.uni-koeln.de} \affiliation{Chennai
Mathematical Institute, Siruseri, 603103, India}
\altaffiliation{Presently at:  Institute for Theoretical Physics, University of Cologne, Germany}

\author{Guillaume Faye}\email{faye@iap.fr}
\affiliation{$\mathcal{G}\mathbb{R}\varepsilon{\mathbb{C}}\mathcal{O}$,
Institut d'Astrophysique de Paris, UMR 7095 CNRS, Sorbonne Universit{\'e}s,
  UPMC Univ Paris 06, F-75014 Paris, France}

\author{K. G. Arun}\email{kgarun@cmi.ac.in} \affiliation{Chennai Mathematical
Institute, Siruseri, 603103, India}

\begin{abstract}

  For black-hole binaries whose spins are (anti-) aligned with respect to the
  orbital angular momentum of the binary, we compute the frequency domain
  phasing coefficients including the quadratic-in-spin terms up to the third
	post-Newtonian (3PN) order, the cubic-in-spin terms at the leading order,
  3.5PN, and the spin-orbit effects up to the 4PN order. In addition, we
  obtain the 2PN spin contributions to the amplitude of the frequency-domain
  gravitational waveforms for non-precessing binaries, using recently derived
  expressions for the time-domain polarization amplitudes of binaries with
  generic spins, complete at that accuracy level. These two results are
  updates to Refs.~\cite{ABFO08} for amplitude and~\cite{WCON2013} for
  phasing. They should be useful to construct banks of templates that model
  accurately non-precessing inspiraling binaries, for parameter estimation
  studies, and or constructing analytical template families that accounts for
  the inspiral-merger-ringdown phases of the binary.

\end{abstract}

\date{\today}
\maketitle

\section{Introduction} \label{sec:intro}

Recently there have been several improvements in modelling spinning binaries
within the post-Newtonian formalism~\cite{Blanchet:2013haa}. These
developments include the computation of relative 2PN spin-orbit (SO) effects
(corresponding to the 3.5PN order) in the equations of
motion~\cite{HS11so,MBFB2012,LS15a} as well as in the precession equations at
the same relative accuracy level, and that of the near-zone metric at the 2PN
order~\cite{BMFB2012}. The work~\cite{BMFB2012} also provided us with the
energy function at 3.5PN order including spin-orbit (linear-in-spins) effects
at the relative 2PN order, which is needed to compute the phase. Further, in
Ref.~\cite{BMB2013}, the 2PN SO contributions were incorporated to the
gravitational-wave energy flux and (time-domain) phasing at the 3.5PN order.
The tail-induced SO corrections to the two latter quantities were investigated
in Ref.~\cite{M3B2013} at the order 4PN, where they are the only spin-orbit
effects. On the other hand, the spin-spin (quadratic-in-spins, SS)
interactions were recently included at the 3PN order~\cite{BFMP2015}, which
means 1PN order beyond the leading SS terms presented in Ref.~\cite{ABFO08}.
In addition, the leading cubic-in-spin terms entering the energy and the
energy flux at 3.5PN were computed in~\cite{Marsat2014}. The 2PN polarizations
$h_{+,\times}$ accounting for both the spin-orbit and spin-spin effects were
calculated explicitly in Ref.~\cite{BFH2012}, extending the earlier works of
Refs.~\cite{K95,WWi96,ABFO08}. Note that the tail-type spin-orbit corrections
entering the 3PN amplitude are also available~\cite{BBF11}. Hence, all spin
contributions to the GW polarizations in the time-domain are known with 2PN
accuracy, while the time-domain phasing is known to the 4PN, 3PN and 3.5PN
orders, for the SO, SS and SSS effects, respectively.

Frequency domain amplitudes for non-precessing binaries, with spins
(anti-)aligned to the orbital angular momentum vector, were first displayed to
the 2PN order in Ref.~\cite{ABFO08}. Their expression complements that of the
3PN accurate polarizations for non-spinning binaries derived
in~\cite{ABIQ04,BFIS08}. They model the spin-orbit effects at the leading
(1.5PN) order and {\it partial} spin-spin effects at the 2PN order. More
precisely, the spin-spin contributions to the GW amplitude presented in
Ref.~\cite{ABFO08} are only those that arise due to couplings involving
\emph{both} spins, \emph{i.e.} of the type (Spin(1)-Spin(2)), as at that time
self-spin corrections (Spin(1)-Spin(1) and (Spin(2)-Spin(2)) were not
available. In this work we make use of the above mentioned recent time-domain
results for GW polarizations with all possible spin-dependent interactions to
construct their frequency domain counterpart complete up to the 2PN order, by
including the new 2PN SO and SS effects (besides those already present
in~\cite{ABFO08}). Frequency-domain phasing with all SO contributions up to
the 3.5PN order --- except for those produced by the black-hole absorption at
the 2.5PN order --- and all SS contributions at the 2PN order, was provided in
Ref.~\cite{WCON2013}. We extend that result by adding the tail-induced
spin-orbit effects at the 4PN order, as well as the quadratic and cubic spin
terms contributing to the phase at the 3PN and 3.5PN orders, respectively.

This paper is organized in the following manner. We begin Sec.~\ref{sec:wfFD}
by showing the form of the Fourier domain signal and specifying our notations.
The rest of the section is split into two parts. Section~\ref{subsec:phase}
presents the phasing formula, which includes the spin-orbit contribution at
the 4PN order, the quadratic spin terms at the 3PN order, and the cubic ones
at the 3.5PN order. In Sec.~\ref{subsec:amp} we list our findings,
complementing the outcomes of Ref.~\cite{ABFO08}, related to the frequency
domain amplitude of the waveform for non-precessing binaries in quasi-circular 
orbits. Finally, in Sec.~\ref{sec:conclusion} we summarize
our results and discuss their implications.

\section{Frequency domain waveforms for non-precessing binaries in 
circular orbits} \label{sec:wfFD}

Since we view this report as an extension of~\cite{ABFO08}, we basically
follow the definitions and notations provided in there. The reader must
refer to that work for details. Nonetheless, we shall provide below some
minimal compendium both to ensure a natural flow in the paper and to
facilitate the reading. The frequency domain amplitude of a signal
$h_\text{strain}$ produced by a gravitational wave $h_{ij}$ can be written,
truncated at some accuracy level, in the following way (see Sec.~VI B of
Ref.~\cite{ABFO08} for a derivation), using geometrical units where $G=c=1$:
\begin{equation}
\tilde{h}_\text{strain}(f)=\frac{M^2}{D_L}
\sqrt{\frac{5\,\pi}{48}}\sum_{n=0}^{4}\sum_{k=1}^{6}
V_k^{n-7/2}\,C_{k}^{(n)}\,e^{\ui (k\,\Psi_\mathrm{SPA}(f/k)-\pi/4)} \, .
\label{eq:hf} \end{equation}

Here, $\tilde{h}_\text{strain}(f)$ denotes the waveform in the frequency
domain\footnote{For the Fourier transform, we adopt the convention that
  $\displaystyle \tilde{h}(f)= \int \ud t \, e^{2\pi \ui \, f t} h(t) $.} as
observed by the detector while $M$ and $D_L$ stand for the total mass and the
luminosity distance of the source, respectively. The index $n$ indicates the
PN order, whereas the index $k$ keeps track of the different harmonics of the
orbital phase. Hence, the above waveform is 2PN accurate and consists of 6
harmonics. For the $k^\mathrm{th}$ harmonic, the PN parameter $v\equiv v(t)$
entering the time domain waveform has been replaced by a function $V_k$ of the
GW frequency $f$, defined as $V_k(f)=(2\,\pi\,M\,f/k)^{1/3}$. The function
$\Psi_\mathrm{SPA}(f)$ represents essentially the phase of the first harmonic
in the frequency domain as obtained under the Stationary Phase Approximation
(SPA)~\cite{ChrisAnand06,ChrisAnand06b} (see Sec.~VI B of~\cite{ABFO08} for
details). Finally, the coefficients $\mathcal{C}_k^{(n)}$'s depend on the
intrinsic parameters of the binary, such as the masses and the spins, as well
as the angular parameters specifying the binary's location and orientation.

The results of the present paper, along with those of Ref.~\cite{ABFO08}, will
allow one to write amplitude corrections completed up to 2PN order with all
possible spin effects. As already stated, the waveform provided
in~\cite{ABFO08} contains terms describing the spin-orbit effects at the
leading order (1.5PN) and \emph{part of} the spin-spin effects (corresponding
to Spin(1)-Spin(2) interactions) at the 2PN order. The coefficients
$\mathcal{C}_k^{(n)}$ through which they appear are explicitly listed in
Appendix~D of Ref.~\cite{ABFO08}. Thus, for the brevity of presentation and
the sake of avoiding repetition, we shall only show here those
$\mathcal{C}_k^{(n)}$'s that get modified due to inclusion of the spin-orbit
and spin-spin effects at the 2PN order, as discussed in Sec.~\ref{sec:intro}.
Below, we shall display our expression for the GW phase and amplitude in two
separate subsections.

\subsection{Corrections to the phasing formula} \label{subsec:phase}

In order to determine the frequency domain phasing we follow the prescription
of Ref.~\cite{DIS02}, which is based on an energy balance argument. In the
case of quasi-circular non-precessing orbits, the two inputs needed for the
phase derivation are the time domain center-of-mass energy $E$ and the energy
flux $\mathcal{F}$ of the binary, both given in terms of the orbital
frequency, the two relations are invariant for a large class of gauge
transformations.

Schematically, we can write for the energy
\begin{eqnarray} E &=& -\frac{\eta m}{2}v^2\left[E_\mathrm{NS}+E_\mathrm{SO}
+E_\mathrm{SS} + E_\mathrm{SSS} \right] \, , \end{eqnarray}
where $E_\mathrm{NS}, E_\mathrm{SO}$, $E_\mathrm{SS}$ and $E_\mathrm{SSS}$
denote the non-spinning, the spin-orbit (linear-in-spins), the spin-spin
(quadratic-in-spin), and the spin-spin-spin (cubic-in-spin) contributions to
the energy, while $\eta=m_1\,m_2/M^2$ represents the symmetric mass ratio
parameter, with $m_1$ and $m_2$ being the masses of the two companions. The
non-spinning part of the energy is currently available to the 4PN accuracy
beyond the Newtonian order~\cite{Damour:2014jta}. However, for the present
purpose, the 3PN expression of Ref.~\cite{BFIJ02}, completed with the results
of~\cite{BDEI04}, is sufficient since there cannot be any 3.5PN terms in the
energy for quasi-circular orbits (see~\cite{Blanchet:2013haa} for a
discussion). The spin-orbit (linear-in-spin) corrections to the conservative
part of the dynamics, starting from the 1.5PN order, are known with a relative
2PN accuracy, i.e., at the 3.5PN order beyond the Newtonian
level~\cite{HS11so,BMFB2012,LS15a}. The same relative accuracy has been
achieved for the spin-spin (quadratic-in-spin) corrections~\cite{HS11s1s2,
  Levi2010, LS15b}, even though it corresponds now to the 4PN order, as the
leading terms of that type arise at the 2PN approximation~\cite{ABFO08}.
However, since the energy flux has not been determined yet with such
precision, it will be sufficient for us to use the spin-spin part of the
energy at the 3PN order. The explicit expressions of the 3.5PN spin-orbit and
the 3PN spin-spin pieces of the energy can be found in the
works~\cite{BMFB2012} and~\cite{BFMP2015}, respectively. As for the
cubic-in-spin pieces, which contribute at the 3.5PN order, they were only
computed recently~\cite{Marsat2014}.

Similarly, the energy flux has the following structure:
\begin{equation} \mathcal{F}=\frac{32}{5}\,\eta^2\,v^{10}
\left[\mathcal{F}_\mathrm{NS} + \mathcal{F}_\mathrm{SO} +
\mathcal{F}_\mathrm{SS} + \mathcal{F}_\mathrm{SSS}\right], 
\end{equation}
where $\mathcal{F}_\mathrm{NS}$, $\mathcal{F}_\mathrm{SO}$,
$\mathcal{F}_\mathrm{SS}$, and $\mathcal{F}_\mathrm{SSS}$ again denote the
non-spinning, spin-orbit, spin-spin, and spin-spin-spin contributions to 
the energy flux. The
non-spinning contributions up to the 3.5PN order beyond the leading
quadrupolar flux are given in Refs.~\cite{BIJ02,BDEI04}. For the spin-orbit
terms, which first appear at the 1.5PN approximation, our current knowledge
extends up to the 4PN order~\cite{M3B2013}. Let us point out that the 4PN
spin-orbit piece of the energy flux comes from the so-called tail effect at
the next-to-leading order (ignoring non spin-orbit terms). This non-linear
effect can be understood as due to the back scattering of the wave on the
spacetime curvature. It is hereditary in nature, which means that it depends
on the past history of the binary evolution. Note that terms of this type (at
the 3PN and 4PN order) are absent from the energy~\cite{M3B2013}. Spin-spin
(or quadratic-in-spin) corrections, starting from the 2PN order, can be found
up to the 3PN order in Refs.~\cite{ABFO08, BFMP2015}. Finally, the
cubic-in-spin terms at the leading 3.5PN approximation were derived
in~\cite{Marsat2014}.

With these time-domain expressions of the energy and the energy flux in hands,
we are in the position to write the frequency domain phasing entailed by the
SPA. Like the expressions above, it has the following general structure:
\begin{equation} \Psi_\mathrm{SPA}(f) = 2\pi f t_\mathrm{c} - \phi_\mathrm{c} +
\left\{\frac{3}{128\eta\,v^{5}}
\left[\psi_\mathrm{NS}+\psi_\mathrm{SO}+\psi_\mathrm{SS}
+\psi_\mathrm{SSS}\right]\right\}_{v=V_1(f)},  \end{equation}
where $\phi_\mathrm{c}$ denote the orbital phase at the instant $t_\mathrm{c}$
of coalescence.\\

The complete 3.5PN accurate frequency domain phasing for non-spinning binaries
is presented in Refs.~\cite{DIS02,AISS05} while the spin-orbit terms up to the
3.5PN accuracy level and the spin-spin terms at the 2PN order are given in
Refs.~\cite{ABFO08,WCON2013}. The contributions to the phasing we add here
include: (i) the tail-induced 4PN spin-orbits terms, (ii) the 3PN
quadratic-in-spin terms, and (iii) the 3.5PN cubic-in-spin terms. Thus, the
spin contributions to the phasing formula may be expressed as
\begin{equation} \psi_\text{Spin}\equiv\psi_\mathrm{SO}+\psi_\mathrm{SS}
+\psi_\mathrm{SSS} ={v^3}\left[\mathcal{P}_3+\mathcal{P}_4\,v+\mathcal{P}_5\,v^2
+\mathcal{P}_6\,v^{3}+\mathcal{P}_7 v^4+\mathcal{P}_8 v^{5} +\cdots\right].
\label{eq:psi_so} \end{equation}

Refs.~\cite{ABFO08,WCON2013} list the explicit expressions for
$\mathcal{P}_3$, $\mathcal{P}_4$ and $\mathcal{P}_5$ with the required accuracies. By
contrast, the coefficients $\mathcal{P}_6$ and $\mathcal{P}_7$ there only
include relative 1.5PN (leading linear-in-spin tail) and relative 2PN
linear-in-spin contributions, respectively. In the present work, as discussed above, we add
the relative 1PN quadratic-in-spin and the leading order cubic-in-spin
corrections. In addition, we introduce a new coefficient
$\mathcal{P}_8$ of order 4PN that corresponds to the tail-induced SO effect.
The modified coefficients $\mathcal{P}_6$, $\mathcal{P}_7$,
and the new coefficient $\mathcal{P}_8$ take the final following form:

\begin{subequations}
\begin{eqnarray}
{\cal P}_6 &=& \pi\,\Big[\frac{2270}{3}\,\delta
\,\boldsymbol{\chi}_\mathrm{a}\cdot\hat{\boldsymbol{\mathrm L}}_\mathrm{N}
+\left(\frac{2270}{3}-520\,\eta\right)
\boldsymbol{\chi}_\mathrm{s}\cdot\hat{\boldsymbol{\mathrm L}}_\mathrm{N}\Big]
+\left(\frac{75515}{144}-\frac{8225}{18}\eta\right)\delta\,
(\boldsymbol{\chi}_\mathrm{a}\cdot\hat{\boldsymbol{\mathrm L}}_\mathrm{N})\, 
(\boldsymbol{\chi}_\mathrm{s}\cdot\hat{\boldsymbol{\mathrm L}}_\mathrm{N})
\nonumber\\&
+&\left(\frac{75515}{288}-\frac{263245}{252}\eta-480\,\eta ^2\right)
(\boldsymbol{\chi}_\mathrm{a}\cdot\hat{\boldsymbol{\mathrm L}}_\mathrm{N})^2
+\left(\frac{75515}{288}-\frac{232415}{504}\eta+\frac{1255}{9}\eta^2\right) 
(\boldsymbol{\chi}_\mathrm{s}\cdot\hat{\boldsymbol{\mathrm L}}_\mathrm{N})^2,\\
{\cal P}_7 &=& 
\left(-\frac{25150083775 }{3048192}
+\frac{26804935}{6048}\eta-\frac{1985}{48}\eta^2\right)
\delta\,\boldsymbol{\chi}_\mathrm{a}\cdot
\hat{\boldsymbol{\mathrm L}}_\mathrm{N}\nonumber\\
&+&\left(-\frac{25150083775}{3048192}+\frac{10566655595}{762048}\eta 
-\frac{1042165}{3024}\eta^2+\frac{5345}{36}\eta^3\right) 
\boldsymbol{\chi}_\mathrm{s}\cdot\hat{\boldsymbol{\mathrm L}}_\mathrm{N}\nonumber\\
&+&\left(\frac{14585 }{24}-2380\,\eta\right)\delta\, 
(\boldsymbol{\chi}_\mathrm{a}\cdot\hat{\boldsymbol{\mathrm L}}_\mathrm{N})^3+
\left(\frac{14585}{24}-\frac{475}{6}\eta +\frac{100}{3}\eta^2\right) 
(\boldsymbol{\chi}_\mathrm{s}\cdot
\hat{\boldsymbol{\mathrm L}}_\mathrm{N})^3\nonumber\\
&+&\left(\frac{14585}{8}-\frac{215}{2}\eta\right) 
\delta\,(\boldsymbol{\chi}_\mathrm{a}\cdot\hat{\boldsymbol{\mathrm L}}_\mathrm{N}) 
(\boldsymbol{\chi}_\mathrm{s}\cdot\hat{\boldsymbol{\mathrm L}}_\mathrm{N})^2+
\left(\frac{14585}{8}-7270\,\eta +80\,\eta ^2\right) 
(\boldsymbol{\chi}_\mathrm{a}\cdot\hat{\boldsymbol{\mathrm L}}_\mathrm{N})^2 
(\boldsymbol{\chi}_\mathrm{s}\cdot\hat{\boldsymbol{\mathrm L}}_\mathrm{N}),\\
{\cal P}_8&=& \pi\,\left[\left(\frac{233915}{168}-
\frac{99185}{252}\eta \right) 
\delta\,\boldsymbol{\chi}_\mathrm{a}\cdot\hat{\boldsymbol{\mathrm L}}_\mathrm{N}+
\left(\frac{233915}{168}-\frac{3970375}{2268}\eta+\frac{19655}{189}\eta^2\right) 
\boldsymbol{\chi}_\mathrm{s}\cdot
\hat{\boldsymbol{\mathrm L}}_\mathrm{N}\right]\left(1-3\ln v\right).
\end{eqnarray}
\end{subequations}

In the above, $\boldsymbol{\chi}_\mathrm{s}$ and
$\boldsymbol{\chi}_\mathrm{a}$ represent symmetric and anti-symmetric
combinations of the spin vectors associated with the binary individual
components $\boldsymbol{\chi}_1$ and $\boldsymbol{\chi}_2$, namely
\begin{eqnarray} \boldsymbol{\chi}_\mathrm{s}&=&\frac{1}{2}(\boldsymbol{\chi}_1
+\boldsymbol{\chi}_2),\nonumber\\
\boldsymbol{\chi}_\mathrm{a}&=&\frac{1}{2}(\boldsymbol{\chi}_1
-\boldsymbol{\chi}_2). \label{chi} \end{eqnarray}
The quantity $\hat{\boldsymbol{\mathrm L}}_\mathrm{N}$ is the unit vector
pointing along the Newtonian orbital angular momentum. Coordinate frames and
parameter conventions used here are identical to the ones employed in
Ref.~\cite{ABFO08}; more details can be found in Sec~II there; the parameter
$\delta=(m_1-m_2)/m$ represents the difference mass ratio. It should be
emphasized that this result completes the SO phasing at the 4PN (relative 2.5PN)
 order, the SS phasing to the 3PN (relative 1PN) order, and the SSS phasing
to the (leading) 3.5PN order in the frequency domain. In order to get the full
4PN phase, ignoring at this stage possible absorption effects for black holes,
one still needs to add: (i) the 4PN non-spinning terms, which would require to
know the energy flux at the same accuracy level, and (ii) the 3.5PN and 4PN SS
terms, of tail and instantaneous types, respectively. The full phasing formula
including the contributions listed in previous works \cite{ABFO08, WCON2013} is being 
provided in a separate file ({\bf supl-mkaf16.m}), both for completeness and for 
convenient use, and is readable in MATHEMATICA.

\subsection{Corrections to the Amplitude: 2PN spin-orbit and spin-spin effects}
\label{subsec:amp}

In this section, we present our findings concerning the amplitude of the
signal from non-precessing binaries. The general structure of the waveform is
given by Eq.~(\ref{eq:hf}). The frequency domain amplitudes in the absence of
spins up to the 2.5PN order, the spin-orbit terms at the 1.5PN order, and
partial spin-spin terms contributing at the 2PN order are listed in
Ref.~\cite{ABFO08}. The related coefficients $\mathcal{C}_{k}^{(n)}$ entering
Eq.~(\ref{eq:hf}) are defined in Eq.(6.13) and (6.14) of \cite{ABFO08} and
have been listed in Appendix~D there. As discussed above, we shall only
provide explicit expressions for those $\mathcal{C}_{k}^{(n)}$'s that get
modified due to inclusion of 2PN spin-orbit and spin-spin terms computed in
the time-domain by Ref.~\cite{BFH2012}. They read:
\begin{subequations} \begin{eqnarray} \label{eq:C14} \mathcal{C}_1^{(4)}&=&s_i
\Biggl\{F_+\Bigg[\delta\,\biggl[\frac{11 \ui}{40}+\frac{5\,\pi}{8}
+\frac{5\ui}{4}\log2 +\left(\frac{7
\ui}{40}+\frac{\pi}{8}+\frac{\ui}{4}\,\log2\right)c_i^2\biggr]
+\delta\,\boldsymbol{\chi}_\mathrm{s}\cdot \hat{\boldsymbol{\mathrm L}}_\mathrm{N}
\biggl[-\frac{711}{448}+\frac{33}{16}\eta
+\left(-\frac{65}{192}-\frac{23}{48}\eta\right)c_i^2\biggr]\nonumber\\&
+&\boldsymbol{\chi}_\mathrm{a}\cdot \hat{\boldsymbol{\mathrm L}}_\mathrm{N}
\biggl[-\frac{711}{448} +\frac{173}{48}\eta
+\left(-\frac{65}{192}+\frac{83}{48}\eta\right) c_i^2\biggr]\Bigg]
+\ui\,c_i\,F_{\times} \Bigg[\delta\,\biggl[\frac{9 \ui}{20} +\frac{3 \pi}{4}+\frac{3
\ui}{2}\log2\biggr]\nonumber\\ &+&\delta\,\boldsymbol{\chi}_\mathrm{s}\cdot
\hat{\boldsymbol{\mathrm L}}_\mathrm{N} \biggl[
\left(-\frac{647}{336}+\frac{41}{24}\eta\right)- \frac{\eta}{8}c_i^2
\biggr]+ \boldsymbol{\chi}_\mathrm{a}\cdot \hat{\boldsymbol{\mathrm L}}_\mathrm{N}
\biggl[ \left(-\frac{647}{336}+\frac{125}{24}\eta\right) +\frac{\eta}{8}
c_i^2\biggr]\Bigg]\Biggr\}\, \Theta(F_\mathrm{cut} - f) \, ,\\ \label{eq:C24}
\mathcal{C}_2^{(4)}&=& \frac{1}{\sqrt{2}}
\Biggl\{F_+\Bigg[\frac{113419241}{40642560}+\frac{152987}{16128}\eta
-\frac{11099}{1152}\eta^2 +\left(\frac{165194153}{40642560}
-\frac{149}{1792}\eta+\frac{6709}{1152}\eta ^2\right)c_i^2 \nonumber\\
&+&\left(\frac{1693}{2016}-\frac{5723}{2016}\eta
+\frac{13}{12}\eta^2\right)c_i^4
-\left(\frac{1}{24}-\frac{5}{24}\eta+\frac{5}{24}\eta ^2\right) c_i^6
+(1+c_i^2)\biggl[
\frac{49}{16}\delta\,(\boldsymbol{\chi}_\mathrm{a}\cdot
\hat{\boldsymbol{\mathrm L}}_\mathrm{N})(\boldsymbol{\chi}_\mathrm{s} \cdot
\hat{\boldsymbol{\mathrm L}}_\mathrm{N}) \nonumber\\
&+&(\boldsymbol{\chi}_\mathrm{a}\cdot \hat{\boldsymbol{\mathrm L}}_\mathrm{N})^2
\left(\frac{49}{32}-6\,\eta\right)
+(\boldsymbol{\chi}_\mathrm{s}\cdot \hat{\boldsymbol{\mathrm L}}_\mathrm{N})^2
\left(\frac{49}{32}-\frac{\eta}{8}\right)
\biggr]\Bigg]
+\ui\,c_i\,F_{\times}\Bigg[\frac{114020009}{20321280}
+\frac{133411}{8064}\eta-\frac{7499}{576}\eta^2
\nonumber \\ &+&\left(\boldsymbol{\chi}_\mathrm{a}\cdot
\hat{\boldsymbol{\mathrm L}}_\mathrm{N}\right){}^2 \left(\frac{49}{16}-12
\eta\right)
+\frac{49}{8}
\delta\,(\boldsymbol{\chi}_\mathrm{a}\cdot
\hat{\boldsymbol{\mathrm L}}_\mathrm{N})(\boldsymbol{\chi}_\mathrm{s}\cdot
\hat{\boldsymbol{\mathrm L}}_\mathrm{N}) +\left(\boldsymbol{\chi}_\mathrm{s}\cdot
\hat{\boldsymbol{\mathrm L}}_\mathrm{N}\right){}^2
\left(\frac{49}{16}-\frac{\eta}{4}\right)
\nonumber\\
&+&\left(\frac{5777}{2520}-\frac{5555}{504}\eta+\frac{34}{3}\eta^2\right)c_i^2
+\left(-\frac{1}{4}+\frac{5}{4}\eta-\frac{5}{4}\eta^2\right)
c_i^4\Bigg]\Biggr\}\, \Theta(2 F_\mathrm{cut} - f) \, ,\\ \label{eq:C34}
\mathcal{C}_3^{(4)}&=&\frac{s_i}{\sqrt{3}} \Biggl\{F_+
\Bigg[\boldsymbol{\chi}_\mathrm{a}\cdot\hat{\boldsymbol{\mathrm L}}_\mathrm{N}
\biggl[\frac{195}{64}-\frac{141}{16}\eta
+\left(\frac{195}{64}-\frac{249}{16}\eta\right) c_i^2\biggr]
+\delta\,\boldsymbol{\chi}_\mathrm{s}\cdot\hat{\boldsymbol{\mathrm L}}_\mathrm{N}
\biggl[\frac{195}{64}-\frac{39}{16}\eta
+\left(\frac{195}{64}+\frac{69}{16}\eta\right)c_i^2\biggr]\nonumber\\
&+&\delta\,(1+c_i^2)\left(-\frac{189 \ui}{40} +\frac{9 \pi}{8}+\frac{27}{4}
\ui\log\left({3\over2}\right)\right)\Bigg] +\ui\,c_i\,F_{\times}
\Bigg[\delta\,\left(-\frac{189 \ui}{20}+\frac{9 \pi}{4} +\frac{27}{2}
\ui\log\left({3\over2}\right)\right)\nonumber\\& +&
\boldsymbol{\chi}_\mathrm{a}\cdot\hat{\boldsymbol{\mathrm L}}_\mathrm{N}
\biggl[\left(\frac{195}{32}-21 \eta \right)-\frac{27}{8}\eta\,c_i^2\biggr]+\delta\,
\boldsymbol{\chi}_\mathrm{s}\cdot\hat{\boldsymbol{\mathrm L}}_N
\biggl[\left(\frac{195}{32}-\frac{3 \eta}{2}\right)+\frac{27}{8} \eta\,
c_i^2\biggr]\Bigg]\Biggr\}\, \Theta(3 F_\mathrm{cut} - f) \, .  \end{eqnarray}
\end{subequations}

Note that in deriving the 2PN terms in the SPA amplitude, we have taken into
account all the spin contributions at the 2PN order instead of the partial
ones that Ref.~\cite{ABFO08} used to be consistent with their spin inputs. To
be more precise, we have resorted to the full expression of $\sigma$ displayed
in Eq.~(6.24) when calculating the quantity ${\cal S}_4$ given by Eq.~(6.11)
of~\cite{ABFO08}. Similar to the phase we also provide a complete list of 
$\mathcal{C}_{k}^{(n)}$'s contributing at the 2PN order in the file 
({\bf supl-mkaf16.m}).

\section{Fourier Transform of the GW modes}

In this section, we provide the GW modes ($h_{\ell m}$) contributing to the
waveform at the 2PN order. For this purpose, we must associate spherical
coordinates $(R, \theta, \phi)$ to the source in such a way that, following
the conventions of~\cite{ABFO08}, $\phi$ vanishes for an observer located on
the earth while $\theta$ coincides with the inclination angle $\iota$ as
measured by the same observer. As usual, the three vectors forming the
standard orthogonal basis are referred to as $e^i_r$, $e^i_\theta$ and
$e^i_\phi$. The complex polarization $h \equiv h_{+}-\ui\,h_{\times}\equiv
-m^i m^j h_{ij}$, with $m^i= e^i_\theta - \ui\, e^i_\phi$, can be conveniently
expanded in terms of the spherical harmonics with spin weight $-2$, the
${}_{-2}Y_{\ell m}(\theta,\phi)$'s, whose precise definition is given by
Eqs.~(4.2)--(4.3) of Ref.~\cite{ABFO08}:
\begin{equation} \label{eq:mode_expansion} h(\theta,\phi) =
\sum_{\ell=2}^{+\infty}\sum_{m=-\ell}^{\ell} h_{\ell m}\;{}_{-2}Y_{\ell
m}(\theta,\phi). \end{equation} 
The $h_{\ell m}$ modes of GW polarization have the following
structure~\cite{BFIS08,ABFO08}:
\begin{equation} h_{\ell m}=\frac{2 \, M \,\eta}{D_L}\;v^2
\sqrt{\frac{16\pi}{5}}\;\hat{h}_{\ell m}\,e^{-\ui m\psi}\, .  
\end{equation}
Those for non-spinning binaries are listed in Eq.~(9.4) of~\cite{BFIS08}
whereas the $\tilde{h}_{\ell m}$'s for spinning binaries can be found in
\cite{ABFO08,BFH2012}. Fourier transforms of these individual modes (as
opposed to that of the full time-domain waveform) may be useful in many
studies at the interface of analytical and numerical relativity. Hence we
systematically provide them below. The procedure for computing those Fourier
transforms is similar to the one used by
Ref.~\cite{ChrisAnand06,ChrisAnand06b} which applied the stationary phase
approximation to the individual harmonics. Following the same procedure, we
obtain the Fourier transforms of the $\tilde{h}_{\ell m}$'s that are relevant
for us. They have the form
\begin{equation} \tilde{h}_{\ell m}(f)=\frac{M^2}{D_L}\,\pi\,
\sqrt{\frac{2 \eta}{3}}\;V_m^{-7/2}\;e^{-\ui (m\,\Psi_\mathrm{SPA}(V_m)+\pi/4)}
\;{\hat{H}_{lm}}(V_m) \, .
\end{equation}
Our results for $\hat{H}_{\ell m}(V_m)\equiv\hat{H}_{\ell m}$, consistently
accounting for all spin effects (as well as for those in absence of spins) up
to the 2PN order, read
\begin{subequations} \begin{eqnarray} \hat{H}_{22} &=&
-1+\left(\frac{323}{224}-\frac{451 \eta}{168}\right) V_2^2 +\Bigg[-\frac{27}{8}
\delta\,  \boldsymbol{\chi}_\mathrm{a}\cdot\hat{\boldsymbol{\mathrm L}}_\mathrm{N}
+\boldsymbol{\chi}_\mathrm{s}\cdot\hat{\boldsymbol{\mathrm L}}_\mathrm{N}
\left(-\frac{27}{8}+\frac{11}{6}\eta\right)\Bigg] V_2^3
+\Bigg[\frac{27312085}{8128512} +\frac{1975055}{338688}\eta \nonumber\\ &-&
\frac{105271}{24192}\eta^2
+\left( \boldsymbol{\chi}_\mathrm{a}\cdot\hat{\boldsymbol{\mathrm L}}_N \right){}^2
\left(\frac{113}{32}-14 \eta \right) +\frac{113}{16} \delta
\,(\boldsymbol{\chi}_\mathrm{a}\cdot \hat{\boldsymbol{\mathrm L}}_\mathrm{N})
(\boldsymbol{\chi}_\mathrm{s}\cdot\hat{\boldsymbol{\mathrm L}}_\mathrm{N}) +\left(
\boldsymbol{\chi}_\mathrm{s}\cdot \hat{\boldsymbol{\mathrm L}}_\mathrm{N}\right ){}^2
\left(\frac{113}{32}-\frac{\eta}{8}\right)
\Bigg]V_2^4 \nonumber\\&+& \mathcal{O}(5) \, , \label{eq:H22}\\
\hat{H}_{21} &=& -{\sqrt{2}\over3}\biggl\{\delta  V_1 -\frac{3}{2}\left(
\boldsymbol{\chi}_\mathrm{a}\cdot \hat{\boldsymbol{\mathrm L}}_\mathrm{N} 
+\delta\,  \boldsymbol{\chi}_\mathrm{s}\cdot
\hat{\boldsymbol{\mathrm L}}_\mathrm{N}\right) V_1^2 +\delta
\left(\frac{335}{672}+\frac{117}{56}\eta\right) V_1^3
+\Bigg[\boldsymbol{\chi}_\mathrm{a}\cdot\hat{\boldsymbol{\mathrm L}}_\mathrm{N}
\left(\frac{4771}{1344}-\frac{11941}{336}\eta\right)\nonumber\\ &+&\delta
\boldsymbol{\chi}_\mathrm{s}\cdot\hat{\boldsymbol{\mathrm L}}_\mathrm{N}
\left(\frac{4771}{1344}-\frac{2549}{336}\eta\right) +\delta
\left(-\frac{\ui}{2}-\pi -2 \ui \log (2)\right)\Bigg] V_1^4\biggr\}
+\mathcal{O}(5)\, ,\label{eq:H21}\\ \hat{H}_{33} &=& -\frac{3}{4}
\sqrt{\frac{5}{7}} \biggl\{\delta  V_3+\delta
\left(-\frac{1945}{672}+\frac{27}{8}\eta\right) V_3^3
+\Bigg[\boldsymbol{\chi}_\mathrm{a}\cdot\hat{\boldsymbol{\mathrm L}}_N
\left(\frac{161}{24}-\frac{85}{3}\eta\right) +\delta \,
\boldsymbol{\chi}_\mathrm{s}\cdot\hat{\boldsymbol{\mathrm L}}_N
\left(\frac{161}{24}-\frac{17}{3}\eta\right)\nonumber\\
&+&\delta\left(-\frac{21 \ui}{5}+\pi +6 \ui \log
\left(\frac{3}{2}\right)\right)\Bigg] V_3^4\biggr\} + \mathcal{O}(5) \,
,\label{eq:H33} \\  \hat{H}_{32} &=& -\frac{1}{3} \sqrt{\frac{5}{7}}
\biggl\{(1-3 \eta ) V_2^2 +4 \eta \,\boldsymbol{\chi}_\mathrm{s}\cdot
\hat{\boldsymbol{\mathrm L}}_\mathrm{N}\, V_2^3
+\left(-\frac{10471}{10080}+\frac{12325}{2016}\eta-\frac{589}{72}\eta^2\right)
V_2^4\biggr\} +\mathcal{O}(5)\, ,\label{eq:H32}\\ \hat{H}_{31} &=& -\frac{1}{{12
\sqrt{7}}}\biggl\{\delta  V_1 +\delta
\left(-\frac{1049}{672}+\frac{17}{24}\eta\right) V_1^3
+\Bigg[\boldsymbol{\chi}_\mathrm{a}\cdot\hat{\boldsymbol{\mathrm L}}_\mathrm{N}
\left(\frac{161}{24}-\frac{73}{3}\eta\right) +\delta\,
\boldsymbol{\chi}_\mathrm{s}\cdot\hat{\boldsymbol{\mathrm L}}_\mathrm{N}
\left(\frac{161}{24}-\frac{29}{3}\eta\right)\nonumber\\ &+&\delta
\left(-\frac{7 \ui}{5}-\pi -2 \ui \log (2)\right)\Bigg] V_1^4\biggr\}
+\mathcal{O}(5)\, ,\label{eq:H31}\\ \hat{H}_{44} &=& -\frac{4}{9}
\sqrt{\frac{10}{7}} \biggl\{(1-3 \eta )\,V_4^2
+\left(-\frac{158383}{36960}+\frac{128221}{7392}\eta-
\frac{1063}{88}\eta^2\right) V_4^4\biggr\}+\mathcal{O}(5)\, ,\label{eq:H44}\\
\hat{H}_{43} &=& -\frac{3}{4} \sqrt{\frac{3}{35}} \biggl\{\delta \,(1-2 \eta )\,
V_3^3+\frac{5}{2}\, \eta\,\left(
\boldsymbol{\chi}_\mathrm{a}\cdot\hat{\boldsymbol{\mathrm L}}_\mathrm{N}
-\delta \,
\boldsymbol{\chi}_\mathrm{s}\cdot\hat{\boldsymbol{\mathrm L}}_\mathrm{N}\right)
V_3^4\biggr\} + \mathcal{O}(5)\, ,\label{eq:H43}\\ \hat{H}_{42} &=&
-\frac{1}{63} \sqrt{5} \biggl\{(1-3 \eta )\, V_2^2
+\left(-\frac{105967}{36960}+\frac{75805}{7392}\eta-
\frac{439}{88}\eta^2\right) V_2^4\biggr\}+\mathcal{O}(5)\, ,\label{eq:H42}\\
\hat{H}_{41} &=& -\frac{1}{84{\sqrt5}}\biggl\{\delta  (1-2 \eta ) V_1^3
+\frac{5}{2}\,\eta\,\left(
\boldsymbol{\chi}_\mathrm{a}\cdot\hat{\boldsymbol{\mathrm L}}_\mathrm{N} -
\delta \,
\boldsymbol{\chi}_\mathrm{s}\cdot\hat{\boldsymbol{\mathrm L}}_\mathrm{N}\right)
V_1^4\biggr\} +\mathcal{O}(5)\, ,\label{eq:H41}\\ \hat{H}_{55} &=&
-\frac{125}{96} \sqrt{\frac{5}{33}} \delta \, (1-2 \eta)\, V_5^3
+\mathcal{O}(5)\, ,\label{eq:H55}\\ \hat{H}_{54} &=& -\frac{16}{9}
\sqrt{\frac{2}{165}} \left(1-5 \eta +5 \eta ^2\right) V_4^4 + \mathcal{O}(5)\,
,\label{eq:H54}\\ \hat{H}_{53} &=& -\frac{9}{32 \sqrt{55}}\,\delta\,  (1-2 \eta
)\, V_3^3 + \mathcal{O}(5)\, ,\label{eq:H53}\\ \hat{ H}_{52} &=& -\frac{2}{27
\sqrt{55}} \left(1-5 \eta +5 \eta^2\right) V_2^4 +\mathcal{O}(5) \,
,\label{eq:H52}\\ \hat{H}_{51} &=& -\frac{1}{144 \sqrt{770}} \delta \, (1-2 \eta
) \,V_1^3 +\mathcal{O}(5) \, ,\label{eq:H51}\\ \hat{H}_{66} &=& -\frac{18}{5}
\sqrt{\frac{3}{143}} \left(1-5 \eta +5 \eta ^2\right) V_6^4 + \mathcal{O}(5)\,
,\label{eq:H66}\\ \hat{H}_{65} &=& \mathcal{O}(5)\, ,\label{eq:H65}\\
\hat{H}_{64} &=& -\frac{128}{495 \sqrt{39}} \left(1-5 \eta +5 \eta ^2\right)
V_4^4 + \mathcal{O}(5) \, ,\label{eq:H64}\\ \hat{H}_{63} &=& \mathcal{O}(5)\,
,\label{eq:H63}\\ \hat{H}_{62} &=& -\frac{2}{297 \sqrt{65}} \left(1-5 \eta +5
\eta ^2\right) V_2^4 + \mathcal{O}(5)\, ,\label{eq:H62}\\ \hat {H}_{61} &=&
\mathcal{O}(5)\, .\label{eq:H61} \end{eqnarray} \end{subequations}

Let us emphasize that the source frame used to express the above polarizations
(and hence the GW modes) is identical to the one of Refs.~\cite{ABFO08,
  BFH2012} (with spin contributions) but differs from that of
Ref.~\cite{BFIS08} (without spinning contributions). The former frame has been
defined so that the azimuthal angle $\phi$ locating the observer vanishes
there, while the latter is such that $\phi=\pi/2$. From
Eq.~\eqref{eq:mode_expansion} and the property of the spin weighted spherical
harmonics, we see that $h_{\ell m}^{(\text{\cite{ABFO08}})} = \ui^m
  h_{\ell m}^{(\text{\cite{BFIS08}})}$. Although we list all the modes contributing 
to the waveform at the 2PN level here, for the convenience of the user, we list these 
expressions in the file ({\bf supl-mkaf16.m}) which we provide as a supplemental material 
to our paper.

\section{Conclusions}
\label{sec:conclusion}

Based on the recent developments in modelling the spinning
binaries~\cite{M3B2013, BFH2012, Marsat2014, BFMP2015}, we have computed the
tail-induced 4PN spin-orbit contribution, the 3PN quadratic-spin correction
and the 3.5PN cubic-spin correction to the frequency domain phasing of the GW
signal, as well as the {\it complete} spin contributions to the amplitude of
the frequency domain waveform at the 2PN order. The 4PN phase presented here
only accounts for tail-induced spin-orbit effects, which must be supplemented
by non-spinning contributions at this order, but these contributions are
currently out of reach due to lack of necessary inputs for the calculation. On
the other hand, some of the higher-order spin effects are still missing beyond
the 3PN order. Those are: (i) the instantaneous quadratic-in-spin
contributions at the 4PN order (including those resulting from the
interactions between the two spins on the one hand, and the effect of the
spin-induced mass quadrupoles of the black holes on the other hand), (ii)
a quadratic-in-spin piece of gravitational-wave tails at the 3.5PN order.
Moreover, when at least one of the two companions is a spinning black hole,
the imprint of the corresponding absorption has yet to be incorporated to the
flux at the 2.5PN order~\cite{Alvi2001, Poisson2004,Porto2008} beyond the
leading quadrupolar piece, with a 1.5PN relative accuracy~\cite{CPY2013}. This
generates additional terms at the 2.5PN, 3.5PN and 4PN orders in the energy
balance equation that is used to determine the orbital phase expression.

Our new frequency domain amplitude corrections involve spin-orbit as well as
spin-spin terms at the 2PN order. The polarizations and the spherical harmonic
modes of the waveform in the frequency domain are now complete at this
approximation level.

These results will be useful for many purposes. One immediate application
would be in the construction of high accuracy templates for the search of
aligned spin binaries~\cite{NitzEtal2013,CantonEtal2014}. The spin effects in
the amplitude and phase of the waveform will also help in reducing the errors
associated with the parameter estimation of the spinning binary
signals~\cite{WCON2013,Favata2013}. In addition, these waveforms could be
useful to study the effect of spins for various tests of strong field gravity
proposed in the literature~\cite{AIQS06a,AIQS06b,MAIS10,YunesPretorius09,LiEtal2011,AgathosEtal2013}. Last but not
least, these terms could play a crucial role in constructing analytical
inspiral-merger-ringdown waveforms~\cite{ajith-etal-IMRspin-PRL2011} including higher GW modes.

\acknowledgments

This work was initiated during the ICTS Program on Numerical relativity
organized by the International Center for Theoretical Sciences, Bangalore,
in June-July 2013. KGA was partly funded by a grant from the Infosys
foundation. Useful conversations with P Ajith and Bala Iyer are gratefully
acknowledged.

\bibliographystyle{apsrev}
\bibliography{ref-list}

\begin{thebibliography}{42}
\expandafter\ifx\csname natexlab\endcsname\relax\def\natexlab#1{#1}\fi
\expandafter\ifx\csname bibnamefont\endcsname\relax
  \def\bibnamefont#1{#1}\fi
\expandafter\ifx\csname bibfnamefont\endcsname\relax
  \def\bibfnamefont#1{#1}\fi
\expandafter\ifx\csname citenamefont\endcsname\relax
  \def\citenamefont#1{#1}\fi
\expandafter\ifx\csname url\endcsname\relax
  \def\url#1{\texttt{#1}}\fi
\expandafter\ifx\csname urlprefix\endcsname\relax\def\urlprefix{URL }\fi
\providecommand{\bibinfo}[2]{#2}
\providecommand{\eprint}[2][]{\url{#2}}

\bibitem[{\citenamefont{Arun et~al.}(2009)\citenamefont{Arun, Buonanno, Faye,
  and Ochsner}}]{ABFO08}
\bibinfo{author}{\bibfnamefont{K.~G.} \bibnamefont{Arun}},
  \bibinfo{author}{\bibfnamefont{A.}~\bibnamefont{Buonanno}},
  \bibinfo{author}{\bibfnamefont{G.}~\bibnamefont{Faye}}, \bibnamefont{and}
  \bibinfo{author}{\bibfnamefont{E.}~\bibnamefont{Ochsner}},
  \bibinfo{journal}{Phys.~Rev.~D} \textbf{\bibinfo{volume}{79}},
  \bibinfo{pages}{104023} (\bibinfo{year}{2009}), \eprint{arXiv:0810.5336}.

\bibitem[{\citenamefont{Wade et~al.}(2013)\citenamefont{Wade, Creighton,
  Ochsner, and Nielsen}}]{WCON2013}
\bibinfo{author}{\bibfnamefont{M.}~\bibnamefont{Wade}},
  \bibinfo{author}{\bibfnamefont{J.~D.~E.} \bibnamefont{Creighton}},
  \bibinfo{author}{\bibfnamefont{E.}~\bibnamefont{Ochsner}}, \bibnamefont{and}
  \bibinfo{author}{\bibfnamefont{A.~B.} \bibnamefont{Nielsen}},
  \bibinfo{journal}{Phys.~Rev.~D} \textbf{\bibinfo{volume}{88}},
  \bibinfo{pages}{083002} (\bibinfo{year}{2013}), \eprint{arXiv:1306.3901}.

\bibitem[{\citenamefont{Blanchet}(2014)}]{Blanchet:2013haa}
\bibinfo{author}{\bibfnamefont{L.}~\bibnamefont{Blanchet}},
  \bibinfo{journal}{Living Reviews in Relativity} \textbf{\bibinfo{volume}{17}},
  \bibinfo{pages}{2} (\bibinfo{year}{2014}), \eprint{arXiv:1310.1528}.

\bibitem[{\citenamefont{{Hartung} and {Steinhoff}}(2011)}]{HS11so}
\bibinfo{author}{\bibfnamefont{J.}~\bibnamefont{{Hartung}}} \bibnamefont{and}
  \bibinfo{author}{\bibfnamefont{J.}~\bibnamefont{{Steinhoff}}},
  \bibinfo{journal}{Annalen der Physik} \textbf{\bibinfo{volume}{523}},
  \bibinfo{pages}{783} (\bibinfo{year}{2011}), \eprint{arXiv:1104.3079},
  \urlprefix\url{http://dx.doi.org/10.1002/andp.201100094}.

\bibitem[{\citenamefont{Marsat et~al.}(2013)\citenamefont{Marsat, Boh{\'e},
  Faye, and Blanchet}}]{MBFB2012}
\bibinfo{author}{\bibfnamefont{S.}~\bibnamefont{Marsat}},
  \bibinfo{author}{\bibfnamefont{A.}~\bibnamefont{Boh{\'e}}},
  \bibinfo{author}{\bibfnamefont{G.}~\bibnamefont{Faye}}, \bibnamefont{and}
  \bibinfo{author}{\bibfnamefont{L.}~\bibnamefont{Blanchet}},
  \bibinfo{journal}{Class.~Quant.~Grav.} \textbf{\bibinfo{volume}{30}},
  \bibinfo{pages}{055007} (\bibinfo{year}{2013}), \eprint{arXiv:1210.4143}.

\bibitem[{\citenamefont{Levi and Steinhoff}(2016{\natexlab{a}})}]{LS15a}
\bibinfo{author}{\bibfnamefont{M.}~\bibnamefont{Levi}} \bibnamefont{and}
  \bibinfo{author}{\bibfnamefont{J.}~\bibnamefont{Steinhoff}},
  \bibinfo{journal}{Journal of Cosmology and Astroparticle Physics}
  \textbf{\bibinfo{volume}{2016}}, \bibinfo{pages}{011}
  (\bibinfo{year}{2016}{\natexlab{a}}),
  \urlprefix\url{http://stacks.iop.org/1475-7516/2016/i=01/a=011}.

\bibitem[{\citenamefont{Boh{\'e}
  et~al.}(2013{\natexlab{a}})\citenamefont{Boh{\'e}, Marsat, Faye, and
  Blanchet}}]{BMFB2012}
\bibinfo{author}{\bibfnamefont{A.}~\bibnamefont{Boh{\'e}}},
  \bibinfo{author}{\bibfnamefont{S.}~\bibnamefont{Marsat}},
  \bibinfo{author}{\bibfnamefont{G.}~\bibnamefont{Faye}}, \bibnamefont{and}
  \bibinfo{author}{\bibfnamefont{L.}~\bibnamefont{Blanchet}},
  \bibinfo{journal}{Class.~Quant.~Grav.} \textbf{\bibinfo{volume}{30}},
  \bibinfo{pages}{075017} (\bibinfo{year}{2013}{\natexlab{a}}),
  \eprint{arXiv:1212.5520}.

\bibitem[{\citenamefont{Boh{\'e}
  et~al.}(2013{\natexlab{b}})\citenamefont{Boh{\'e}, Marsat, and
  Blanchet}}]{BMB2013}
\bibinfo{author}{\bibfnamefont{A.}~\bibnamefont{Boh{\'e}}},
  \bibinfo{author}{\bibfnamefont{S.}~\bibnamefont{Marsat}}, \bibnamefont{and}
  \bibinfo{author}{\bibfnamefont{L.}~\bibnamefont{Blanchet}},
  \bibinfo{journal}{Class.~Quant.~Grav.} \textbf{\bibinfo{volume}{30}},
  \bibinfo{pages}{135009} (\bibinfo{year}{2013}{\natexlab{b}}),
  \eprint{arXiv:1303.7412}.

\bibitem[{\citenamefont{Marsat et~al.}(2014)\citenamefont{Marsat, Boh{\'e},
  Blanchet, and Buonanno}}]{M3B2013}
\bibinfo{author}{\bibfnamefont{S.}~\bibnamefont{Marsat}},
  \bibinfo{author}{\bibfnamefont{A.}~\bibnamefont{Boh{\'e}}},
  \bibinfo{author}{\bibfnamefont{L.}~\bibnamefont{Blanchet}}, \bibnamefont{and}
  \bibinfo{author}{\bibfnamefont{A.}~\bibnamefont{Buonanno}},
  \bibinfo{journal}{Class.~Quant.~Grav.} \textbf{\bibinfo{volume}{31}},
  \bibinfo{pages}{025023} (\bibinfo{year}{2014}), \eprint{arXiv:1307.6793}.

\bibitem[{\citenamefont{{Boh{\'e}} et~al.}(2015)\citenamefont{{Boh{\'e}},
  {Faye}, {Marsat}, and {Porter}}}]{BFMP2015}
\bibinfo{author}{\bibfnamefont{A.}~\bibnamefont{{Boh{\'e}}}},
  \bibinfo{author}{\bibfnamefont{G.}~\bibnamefont{{Faye}}},
  \bibinfo{author}{\bibfnamefont{S.}~\bibnamefont{{Marsat}}}, \bibnamefont{and}
  \bibinfo{author}{\bibfnamefont{E.~K.} \bibnamefont{{Porter}}},
  \bibinfo{journal}{ArXiv e-prints}  (\bibinfo{year}{2015}),
  \eprint{arXiv:1501.01529}.

\bibitem[{\citenamefont{{Marsat}}(2015)}]{Marsat2014}
\bibinfo{author}{\bibfnamefont{S.}~\bibnamefont{{Marsat}}},
  \bibinfo{journal}{Class.~Quant.~Grav.}
  \textbf{\bibinfo{volume}{32}}, \bibinfo{eid}{085008} (\bibinfo{year}{2015}),
  \eprint{arXiv:1411.4118}.

\bibitem[{\citenamefont{Buonanno et~al.}(2013)\citenamefont{Buonanno, Faye, and
  Hinderer}}]{BFH2012}
\bibinfo{author}{\bibfnamefont{A.}~\bibnamefont{Buonanno}},
  \bibinfo{author}{\bibfnamefont{G.}~\bibnamefont{Faye}}, \bibnamefont{and}
  \bibinfo{author}{\bibfnamefont{T.}~\bibnamefont{Hinderer}},
  \bibinfo{journal}{Phys.~Rev.~D} \textbf{\bibinfo{volume}{87}},
  \bibinfo{pages}{044009} (\bibinfo{year}{2013}), \eprint{arXiv:1209.6349}.

\bibitem[{\citenamefont{Kidder}(1995)}]{K95}
\bibinfo{author}{\bibfnamefont{L.}~\bibnamefont{Kidder}},
  \bibinfo{journal}{Phys.~Rev.~D} \textbf{\bibinfo{volume}{52}},
  \bibinfo{pages}{821} (\bibinfo{year}{1995}).

\bibitem[{\citenamefont{Will and Wiseman}(1996)}]{WWi96}
\bibinfo{author}{\bibfnamefont{C.}~\bibnamefont{Will}} \bibnamefont{and}
  \bibinfo{author}{\bibfnamefont{A.}~\bibnamefont{Wiseman}},
  \bibinfo{journal}{Phys.~Rev.~D} \textbf{\bibinfo{volume}{54}},
  \bibinfo{pages}{4813} (\bibinfo{year}{1996}).

\bibitem[{\citenamefont{Blanchet et~al.}(2011)\citenamefont{Blanchet, Buonanno,
  and Faye}}]{BBF11}
\bibinfo{author}{\bibfnamefont{L.}~\bibnamefont{Blanchet}},
  \bibinfo{author}{\bibfnamefont{A.}~\bibnamefont{Buonanno}}, \bibnamefont{and}
  \bibinfo{author}{\bibfnamefont{G.}~\bibnamefont{Faye}},
  \bibinfo{journal}{Phys.~Rev.~D} \textbf{\bibinfo{volume}{84}},
  \bibinfo{pages}{064041} (\bibinfo{year}{2011}),
  \urlprefix\url{http://link.aps.org/doi/10.1103/PhysRevD.84.064041}.

\bibitem[{\citenamefont{Arun et~al.}(2004)\citenamefont{Arun, Blanchet, Iyer,
  and Qusailah}}]{ABIQ04}
\bibinfo{author}{\bibfnamefont{K.~G.} \bibnamefont{Arun}},
  \bibinfo{author}{\bibfnamefont{L.}~\bibnamefont{Blanchet}},
  \bibinfo{author}{\bibfnamefont{B.~R.} \bibnamefont{Iyer}}, \bibnamefont{and}
  \bibinfo{author}{\bibfnamefont{M.~S.~S.} \bibnamefont{Qusailah}},
  \bibinfo{journal}{Class.~Quant.~Grav.} \textbf{\bibinfo{volume}{21}},
  \bibinfo{pages}{3771} (\bibinfo{year}{2004}), \bibinfo{note}{erratum-ibid.
  {\bf 22}, 3115 (2005)}, \eprint{gr-qc/0404185}.

\bibitem[{\citenamefont{Blanchet et~al.}(2008)\citenamefont{Blanchet, Faye,
  Iyer, and Sinha}}]{BFIS08}
\bibinfo{author}{\bibfnamefont{L.}~\bibnamefont{Blanchet}},
  \bibinfo{author}{\bibfnamefont{G.}~\bibnamefont{Faye}},
  \bibinfo{author}{\bibfnamefont{B.~R.} \bibnamefont{Iyer}}, \bibnamefont{and}
  \bibinfo{author}{\bibfnamefont{S.}~\bibnamefont{Sinha}},
  \bibinfo{journal}{Class.~Quant.~Grav.} \textbf{\bibinfo{volume}{25}},
  \bibinfo{pages}{165003} (\bibinfo{year}{2008}), \eprint{arXiv:0802.1249}.

\bibitem[{\citenamefont{Van Den~Broeck and Sengupta}(2007)}]{ChrisAnand06}
\bibinfo{author}{\bibfnamefont{C.}~\bibnamefont{Van Den~Broeck}}
  \bibnamefont{and} \bibinfo{author}{\bibfnamefont{A.}~\bibnamefont{Sengupta}},
  \bibinfo{journal}{Class.~Quant.~Grav.} \textbf{\bibinfo{volume}{24}},
  \bibinfo{pages}{155} (\bibinfo{year}{2007}), \eprint{gr-qc/0607092}.

\bibitem[{\citenamefont{{Van Den Broeck} and {Sengupta}}(2007)}]{ChrisAnand06b}
\bibinfo{author}{\bibfnamefont{C.}~\bibnamefont{{Van Den Broeck}}}
  \bibnamefont{and} \bibinfo{author}{\bibfnamefont{A.~S.}
  \bibnamefont{{Sengupta}}}, \bibinfo{journal}{Class.~Quant.~Grav.}
  \textbf{\bibinfo{volume}{24}}, \bibinfo{pages}{1089} (\bibinfo{year}{2007}),
  \eprint{gr-qc/0610126}.

\bibitem[{\citenamefont{Damour et~al.}(2002)\citenamefont{Damour, Iyer, and
  Sathyaprakash}}]{DIS02}
\bibinfo{author}{\bibfnamefont{T.}~\bibnamefont{Damour}},
  \bibinfo{author}{\bibfnamefont{B.~R.} \bibnamefont{Iyer}}, \bibnamefont{and}
  \bibinfo{author}{\bibfnamefont{B.~S.} \bibnamefont{Sathyaprakash}},
  \bibinfo{journal}{Phys.~Rev.~D} \textbf{\bibinfo{volume}{66}},
  \bibinfo{pages}{027502} (\bibinfo{year}{2002}),
  \bibinfo{note}{erratum-{ibid}~{\bf 66}, 027502 (2002)},
  \eprint{gr-qc/0207021}.

\bibitem[{\citenamefont{Damour et~al.}(2014)\citenamefont{Damour, Jaranowski,
  and Sch\"afer}}]{Damour:2014jta}
\bibinfo{author}{\bibfnamefont{T.}~\bibnamefont{Damour}},
  \bibinfo{author}{\bibfnamefont{P.}~\bibnamefont{Jaranowski}},
  \bibnamefont{and}
  \bibinfo{author}{\bibfnamefont{G.}~\bibnamefont{Sch\"afer}},
  \bibinfo{journal}{Phys.~Rev.~D} \textbf{\bibinfo{volume}{89}},
  \bibinfo{pages}{064058} (\bibinfo{year}{2014}), \eprint{arXiv:1401.4548},
  \urlprefix\url{http://link.aps.org/doi/10.1103/PhysRevD.89.064058}.

\bibitem[{\citenamefont{Blanchet
  et~al.}(2002{\natexlab{a}})\citenamefont{Blanchet, Faye, Iyer, and
  Joguet}}]{BFIJ02}
\bibinfo{author}{\bibfnamefont{L.}~\bibnamefont{Blanchet}},
  \bibinfo{author}{\bibfnamefont{G.}~\bibnamefont{Faye}},
  \bibinfo{author}{\bibfnamefont{B.~R.} \bibnamefont{Iyer}}, \bibnamefont{and}
  \bibinfo{author}{\bibfnamefont{B.}~\bibnamefont{Joguet}},
  \bibinfo{journal}{Phys.~Rev.~D} \textbf{\bibinfo{volume}{65}},
  \bibinfo{pages}{061501(R)} (\bibinfo{year}{2002}{\natexlab{a}}),
  \bibinfo{note}{{Erratum-ibid~{\bf 71}, 129902(E) (2005)}},
  \eprint{gr-qc/0105099}.

\bibitem[{\citenamefont{Blanchet et~al.}(2004)\citenamefont{Blanchet, Damour,
  Esposito-Far{\`e}se, and Iyer}}]{BDEI04}
\bibinfo{author}{\bibfnamefont{L.}~\bibnamefont{Blanchet}},
  \bibinfo{author}{\bibfnamefont{T.}~\bibnamefont{Damour}},
  \bibinfo{author}{\bibfnamefont{G.}~\bibnamefont{Esposito-Far{\`e}se}},
  \bibnamefont{and} \bibinfo{author}{\bibfnamefont{B.~R.} \bibnamefont{Iyer}},
  \bibinfo{journal}{Phys.~Rev.~Lett.} \textbf{\bibinfo{volume}{93}},
  \bibinfo{pages}{091101} (\bibinfo{year}{2004}), \eprint{gr-qc/0406012}.

\bibitem[{\citenamefont{Hartung and Steinhoff}(2011)}]{HS11s1s2}
\bibinfo{author}{\bibfnamefont{J.}~\bibnamefont{Hartung}} \bibnamefont{and}
  \bibinfo{author}{\bibfnamefont{J.}~\bibnamefont{Steinhoff}},
  \bibinfo{journal}{Annalen der Physik} \textbf{\bibinfo{volume}{523}},
  \bibinfo{pages}{919} (\bibinfo{year}{2011}), ISSN \bibinfo{issn}{1521-3889},
  \urlprefix\url{http://dx.doi.org/10.1002/andp.201100163}.

\bibitem[{\citenamefont{Levi}(2010)}]{Levi2010}
\bibinfo{author}{\bibfnamefont{M.}~\bibnamefont{Levi}}, 
\bibinfo{journal}{Phys.~Rev.~D} \textbf{\bibinfo{volume}{82}}, \bibinfo{pages}{064029}
  (\bibinfo{year}{2010}),
  \urlprefix\url{http://link.aps.org/doi/10.1103/PhysRevD.82.064029}.

\bibitem[{\citenamefont{Levi and Steinhoff}(2016{\natexlab{b}})}]{LS15b}
\bibinfo{author}{\bibfnamefont{M.}~\bibnamefont{Levi}} \bibnamefont{and}
  \bibinfo{author}{\bibfnamefont{J.}~\bibnamefont{Steinhoff}},
  \bibinfo{journal}{Journal of Cosmology and Astroparticle Physics}
  \textbf{\bibinfo{volume}{2016}}, \bibinfo{pages}{008}
  (\bibinfo{year}{2016}{\natexlab{b}}),
  \urlprefix\url{http://stacks.iop.org/1475-7516/2016/i=01/a=008}.

\bibitem[{\citenamefont{Blanchet
  et~al.}(2002{\natexlab{b}})\citenamefont{Blanchet, Iyer, and Joguet}}]{BIJ02}
\bibinfo{author}{\bibfnamefont{L.}~\bibnamefont{Blanchet}},
  \bibinfo{author}{\bibfnamefont{B.~R.} \bibnamefont{Iyer}}, \bibnamefont{and}
  \bibinfo{author}{\bibfnamefont{B.}~\bibnamefont{Joguet}},
  \bibinfo{journal}{Phys.~Rev.~D} \textbf{\bibinfo{volume}{65}},
  \bibinfo{pages}{064005} (\bibinfo{year}{2002}{\natexlab{b}}),
  \bibinfo{note}{{Erratum-ibid~{\bf 71}, 129903(E) (2005)}},
  \eprint{gr-qc/0105098}.

\bibitem[{\citenamefont{Arun et~al.}(2005)\citenamefont{Arun, Iyer,
  Sathyaprakash, and Sundararajan}}]{AISS05}
\bibinfo{author}{\bibfnamefont{K.~G.} \bibnamefont{Arun}},
  \bibinfo{author}{\bibfnamefont{B.~R.} \bibnamefont{Iyer}},
  \bibinfo{author}{\bibfnamefont{B.~S.} \bibnamefont{Sathyaprakash}},
  \bibnamefont{and} \bibinfo{author}{\bibfnamefont{P.~A.}
  \bibnamefont{Sundararajan}}, \bibinfo{journal}{Phys.~Rev.~D}
  \textbf{\bibinfo{volume}{71}}, \bibinfo{pages}{084008}
  (\bibinfo{year}{2005}), \bibinfo{note}{erratum-ibid. ~{\bf D } 72, 069903
  (2005)}, \eprint{gr-qc/0411146}.

\bibitem[{\citenamefont{Alvi}(2001)}]{Alvi2001}
\bibinfo{author}{\bibfnamefont{K.}~\bibnamefont{Alvi}}, 
\bibinfo{journal}{Phys.~Rev.~D} \textbf{\bibinfo{volume}{64}}, \bibinfo{pages}{104020}
  (\bibinfo{year}{2001}),
  \urlprefix\url{http://link.aps.org/doi/10.1103/PhysRevD.64.104020}.

\bibitem[{\citenamefont{Poisson}(2004)}]{Poisson2004}
\bibinfo{author}{\bibfnamefont{E.}~\bibnamefont{Poisson}},
  \bibinfo{journal}{Phys.~Rev.~D} \textbf{\bibinfo{volume}{70}},
  \bibinfo{pages}{084044} (\bibinfo{year}{2004}),
  \urlprefix\url{http://link.aps.org/doi/10.1103/PhysRevD.70.084044}.

\bibitem[{\citenamefont{Porto}(2008)}]{Porto2008}
\bibinfo{author}{\bibfnamefont{R.~A.} \bibnamefont{Porto}},
  \bibinfo{journal}{Phys.~Rev.~D} \textbf{\bibinfo{volume}{77}},
  \bibinfo{pages}{064026} (\bibinfo{year}{2008}),
  \urlprefix\url{http://link.aps.org/doi/10.1103/PhysRevD.77.064026}.

\bibitem[{\citenamefont{Chatziioannou et~al.}(2013)\citenamefont{Chatziioannou,
  Poisson, and Yunes}}]{CPY2013}
\bibinfo{author}{\bibfnamefont{K.}~\bibnamefont{Chatziioannou}},
  \bibinfo{author}{\bibfnamefont{E.}~\bibnamefont{Poisson}}, \bibnamefont{and}
  \bibinfo{author}{\bibfnamefont{N.}~\bibnamefont{Yunes}},
  \bibinfo{journal}{Phys.~Rev.~D} \textbf{\bibinfo{volume}{87}},
  \bibinfo{pages}{044022} (\bibinfo{year}{2013}),
  \urlprefix\url{http://link.aps.org/doi/10.1103/PhysRevD.87.044022}.

\bibitem[{\citenamefont{Nitz et~al.}(2013)\citenamefont{Nitz, Lundgren, Brown,
  Ochsner, Keppel et~al.}}]{NitzEtal2013}
\bibinfo{author}{\bibfnamefont{A.~H.} \bibnamefont{Nitz}},
  \bibinfo{author}{\bibfnamefont{A.}~\bibnamefont{Lundgren}},
  \bibinfo{author}{\bibfnamefont{D.~A.} \bibnamefont{Brown}},
  \bibinfo{author}{\bibfnamefont{E.}~\bibnamefont{Ochsner}},
  \bibinfo{author}{\bibfnamefont{D.}~\bibnamefont{Keppel}},
  \bibnamefont{et~al.}, \bibinfo{journal}{Phys.~Rev.~D}
  \textbf{\bibinfo{volume}{88}}, \bibinfo{pages}{124039}
  (\bibinfo{year}{2013}), \eprint{arXiv:1307.1757}.

\bibitem[{\citenamefont{Canton et~al.}(2014)\citenamefont{Canton, Nitz,
  Lundgren, Nielsen, Brown et~al.}}]{CantonEtal2014}
\bibinfo{author}{\bibfnamefont{T.~D.} \bibnamefont{Canton}},
  \bibinfo{author}{\bibfnamefont{A.~H.} \bibnamefont{Nitz}},
  \bibinfo{author}{\bibfnamefont{A.~P.} \bibnamefont{Lundgren}},
  \bibinfo{author}{\bibfnamefont{A.~B.} \bibnamefont{Nielsen}},
  \bibinfo{author}{\bibfnamefont{D.~A.} \bibnamefont{Brown}},
  \bibnamefont{et~al.}, \bibinfo{journal}{Phys.~Rev.~D}
  \textbf{\bibinfo{volume}{90}}, \bibinfo{pages}{082004}
  (\bibinfo{year}{2014}), \eprint{arXiv:1405.6731}.

\bibitem[{\citenamefont{Favata}(2014)}]{Favata2013}
\bibinfo{author}{\bibfnamefont{M.}~\bibnamefont{Favata}},
  \bibinfo{journal}{Phys.~Rev.~Lett.} \textbf{\bibinfo{volume}{112}},
  \bibinfo{pages}{101101} (\bibinfo{year}{2014}), \eprint{arXiv:1310.8288}.

\bibitem[{\citenamefont{Arun et~al.}(2006{\natexlab{a}})\citenamefont{Arun,
  Iyer, Qusailah, and Sathyaprakash}}]{AIQS06a}
\bibinfo{author}{\bibfnamefont{K.~G.} \bibnamefont{Arun}},
  \bibinfo{author}{\bibfnamefont{B.~R.} \bibnamefont{Iyer}},
  \bibinfo{author}{\bibfnamefont{M.~S.~S.} \bibnamefont{Qusailah}},
  \bibnamefont{and} \bibinfo{author}{\bibfnamefont{B.~S.}
  \bibnamefont{Sathyaprakash}}, \bibinfo{journal}{Class.~Quant.~Grav.}
  \textbf{\bibinfo{volume}{23}}, \bibinfo{pages}{L37}
  (\bibinfo{year}{2006}{\natexlab{a}}), \eprint{gr-qc/0604018}.

\bibitem[{\citenamefont{Arun et~al.}(2006{\natexlab{b}})\citenamefont{Arun,
  Iyer, Qusailah, and Sathyaprakash}}]{AIQS06b}
\bibinfo{author}{\bibfnamefont{K.~G.} \bibnamefont{Arun}},
  \bibinfo{author}{\bibfnamefont{B.~R.} \bibnamefont{Iyer}},
  \bibinfo{author}{\bibfnamefont{M.~S.~S.} \bibnamefont{Qusailah}},
  \bibnamefont{and} \bibinfo{author}{\bibfnamefont{B.~S.}
  \bibnamefont{Sathyaprakash}}, \bibinfo{journal}{Phys.~Rev.~D}
  \textbf{\bibinfo{volume}{74}}, \bibinfo{pages}{024006}
  (\bibinfo{year}{2006}{\natexlab{b}}), \eprint{gr-qc/0604067}.

\bibitem[{\citenamefont{Mishra et~al.}(2010)\citenamefont{Mishra, Arun, Iyer,
  and Sathyaprakash}}]{MAIS10}
\bibinfo{author}{\bibfnamefont{C.~K.} \bibnamefont{Mishra}},
  \bibinfo{author}{\bibfnamefont{K.~G.} \bibnamefont{Arun}},
  \bibinfo{author}{\bibfnamefont{B.~R.} \bibnamefont{Iyer}}, \bibnamefont{and}
  \bibinfo{author}{\bibfnamefont{B.~S.} \bibnamefont{Sathyaprakash}},
  \bibinfo{journal}{Phys.~Rev.~D} \textbf{\bibinfo{volume}{82}},
  \bibinfo{pages}{064010} (\bibinfo{year}{2010}), \eprint{arXiv:1005.0304}.

\bibitem[{\citenamefont{Yunes and Pretorius}(2009)}]{YunesPretorius09}
\bibinfo{author}{\bibfnamefont{N.}~\bibnamefont{Yunes}} \bibnamefont{and}
  \bibinfo{author}{\bibfnamefont{F.}~\bibnamefont{Pretorius}},
  \bibinfo{journal}{Phys.~Rev.~D} \textbf{\bibinfo{volume}{80}},
  \bibinfo{pages}{122003} (\bibinfo{year}{2009}), \eprint{arXiv:0909.3328}.

\bibitem[{\citenamefont{Li et~al.}(2012)}]{LiEtal2011}
\bibinfo{author}{\bibfnamefont{T.~G.~F.} \bibnamefont{Li}}
  \bibnamefont{et~al.}, \bibinfo{journal}{Phys.~Rev.~D}
  \textbf{\bibinfo{volume}{85}}, \bibinfo{pages}{082003}
  (\bibinfo{year}{2012}), \eprint{arXiv:1111.5274}.

\bibitem[{\citenamefont{Agathos et~al.}(2015)\citenamefont{Agathos, Del~Pozzo,
  Li, Van Den~Broeck, Veitch, and Vitale}}]{AgathosEtal2013}
\bibinfo{author}{\bibfnamefont{M.}~\bibnamefont{Agathos}},
  \bibinfo{author}{\bibfnamefont{W.}~\bibnamefont{Del~Pozzo}},
  \bibinfo{author}{\bibfnamefont{T.~G.~F.} \bibnamefont{Li}},
  \bibinfo{author}{\bibfnamefont{C.}~\bibnamefont{Van Den~Broeck}},
  \bibinfo{author}{\bibfnamefont{J.}~\bibnamefont{Veitch}}, \bibnamefont{and}
  \bibinfo{author}{\bibfnamefont{S.}~\bibnamefont{Vitale}}, in
  \emph{\bibinfo{booktitle}{{Proceedings, 13th Marcel Grossmann Meeting on
  Recent Developments in Theoretical and Experimental General Relativity,
  Astrophysics, and Relativistic Field Theories (MG13)}}}
  (\bibinfo{year}{2015}), pp. \bibinfo{pages}{1710--1712}, \eprint{arXiv:1305.2963},
  \urlprefix\url{http://inspirehep.net/record/1233350/files/arXiv:1305.2963.pdf}.

\bibitem [{\citenamefont {Ajith et.~al.}(2011)}]{ajith-etal-IMRspin-PRL2011}%
  {\bibinfo {author} {\bibfnamefont {P.}~\bibnamefont
  {{Ajith}}},
  \bibinfo {author} {\bibfnamefont {M.}~\bibnamefont {{Hannam}}},
  \bibinfo {author} {\bibfnamefont {S.}~\bibnamefont {{Husa}}},
  \bibinfo {author} {\bibfnamefont {Y.}~\bibnamefont {{Chen}}},
  \bibinfo {author} {\bibfnamefont {B.}~\bibnamefont {{Br{\"u}gmann}}},
  \bibinfo {author} {\bibfnamefont {N.}~\bibnamefont {{Dorband}}},
  \bibinfo {author} {\bibfnamefont {D.}~\bibnamefont {{M{\"u}ller}}},
  \bibinfo {author} {\bibfnamefont {F.}~\bibnamefont {{Ohme}}},
  \bibinfo {author} {\bibfnamefont {D.}~\bibnamefont {{Pollney}}},
  \bibinfo {author} {\bibfnamefont {C.}~\bibnamefont {{Reisswig}}},
  \bibinfo {author} {\bibfnamefont {L.}~\bibnamefont {{Santamar{\'{\i}}a}}},
  \ and\
  \bibinfo {author} {\bibfnamefont {J.}~\bibnamefont {{Seiler}}},\ }
  {\bibfield  {journal} {\bibinfo  {journal}
  {Phys. Rev. Lett.}\ }\textbf {\bibinfo {volume} {106}},\ \bibinfo
  {eid} {241101} (\bibinfo {year} {2011})}, {arXiv:0909.2867}
\end{thebibliography}

\end{document}